\documentclass[12pt]{iopart}

\usepackage{iopams}  

\usepackage[dvips]{graphicx,color}
\usepackage{delarray}
\usepackage{bm}

\newcommand{\braket}[2]{\langle {#1} | {#2} \rangle}
\newcommand{\ket}[1]{| { #1} \rangle}
\newcommand{\bra}[1]{ \langle {#1}  |}

\begin{document}

\title{Tight bound on coherent-state-based entanglement generation over lossy channels}

\author{Koji Azuma, Naoya Sota, Masato Koashi and Nobuyuki Imoto}

\address{Division of Materials Physics, Department of Materials Engineering Science,
Graduate School of Engineering Science, Osaka University,
1-3 Machikaneyama, Toyonaka, Osaka 560-8531, Japan}
\ead{azuma@qi.mp.es.osaka-u.ac.jp}

\begin{abstract}
The first stage of the hybrid quantum repeaters is entanglement generation based on transmission of pulses in coherent states over a lossy channel.
Protocols to make entanglement with only one type of error are favorable for rendering subsequent entanglement distillation efficient. 
Here we provide the tight upper bound on performances of these protocols that is determined only by the channel loss. In addition, we show that this bound is achievable by utilizing a proposed protocol [quant-ph/0811.3100] composed of a simple combination of linear optical elements and photon-number-resolving detectors. 
\end{abstract}


\maketitle

\section{Introduction}
Quantum communication is the key technique to enable important applications such as quantum teleportation \cite{B93}, quantum key distribution \cite{E91}, and distributed quantum computation \cite{G97}.
A solution to realize arbitrary long-distance quantum communication over a practical transmission channel is to invoke a quantum repeater protocol \cite{B98,D99,D01,B07,Z07,J07, C06,C05,L06,La06,L08,M08,S08}. 
One of the promising candidates is the so-called hybrid quantum repeater protocol \cite{L06,La06,L08,M08,S08}, which features its `hybrid' aspect combining `discrete' states of quantum memories and `continuous' variables of optical pulses in coherent states. 
As an advantage of the hybrid quantum repeater protocols, 
all the stages in the repeater protocol -- entanglement generation, entanglement distillation \cite{B96,D96,B96a}, and entanglement swapping \cite{Z93} -- are shown to be implementable \cite{L06,La06,S06} only by realizing a quantum memory that can interact with optical pulses in the form of
\begin{equation}
\eqalign{ \hat{V} \ket{0}_A \ket{\alpha}_a = \ket{0}_A \ket{\alpha_0}_a, \cr
\hat{V} \ket{1}_A \ket{\alpha}_a = \ket{1}_A \ket{\alpha_1}_a ,}\label{eq:V}
\end{equation}
where $\hat{V}$ is a unitary operator, $\ket{\alpha}_a$ and $\{ \ket{\alpha_j}_a \}_{j=0,1}$ are coherent states of the pulse mode $a$, and $\{ \ket{j}_A \}_{j=0,1}$ are states of the memory.
In the stream of the stages, an undoubted art to achieve higher efficiencies is to find a good entanglement generation protocol leaving the quantum memories in entanglement that is efficiently distillable at the distillation stage.
Until now, there have been many proposals to achieve higher efficiencies in the entanglement generation stage \cite{L06,La06,L08,M08,S08},
and recent results have shown that protocols to make entanglement with only one type of error are favorable for rendering subsequent entanglement distillation efficient \cite{L08,M08}. 

In this paper, considering the protocols that can generate entanglement with only one type of error by transmitting pulses in coherent states through a lossy channel, we provide the tight upper bound on the performances of these protocols stated in terms of the average singlet fraction of generated entanglement and by the success probability.
This bound is determined only by the channel loss, i.e., 
the length of the channel.
In order to derive the bound, we require no additional assumption, differently from Ref. \cite{S08} where the quantum memory of the sender is additionally assumed to start from a symmetric state $(\ket{0}_A + \ket{1}_A)/\sqrt{2}$. 
Our general bound is shown to be achievable by utilizing a proposed protocol \cite{S08} that is realizable by linear optical elements and photon-number-resolving detectors.

This paper is organized as follows. In Sec.~\ref{se:2}, we define protocols to generate entanglement with only one type of error, and the measure of the performance.
We derive an upper bound on those performances in Sec.~\ref{se:3}, 
which is the main theorem in this paper.
In Sec.~\ref{se:4}, we show that the upper bound is achievable by convex combination of the protocol proposed in Ref.~\cite{S08} and a trivial protocol.
In Sec.~\ref{se:5}, we derive an explicit expression of the tight upper bound as a function of the transmittance of the channel loss.
Section~\ref{se:6} concludes the paper.

\section{Single-error-type entanglement generation and the measure of its performance}\label{se:2}

Let us define the family of single-error-type entanglement generation protocols considered in this paper.
We require Alice and Bob to make an entangled state with only one type of error. 
More precisely, Alice and Bob are required to make qubits $AB$ in an entangled state that can be transformed into a state contained in the subspace spanned by Bell states $\{\ket{\Phi^{\pm}}_{AB}\}$ via local unitary operations, where $\ket{\Phi^{\pm}}_{AB}:=(\ket{00}_{AB} \pm \ket{11}_{AB})/\sqrt{2}$.

\begin{figure}[b]
  \begin{center}
    \includegraphics[keepaspectratio=true,height=80mm]{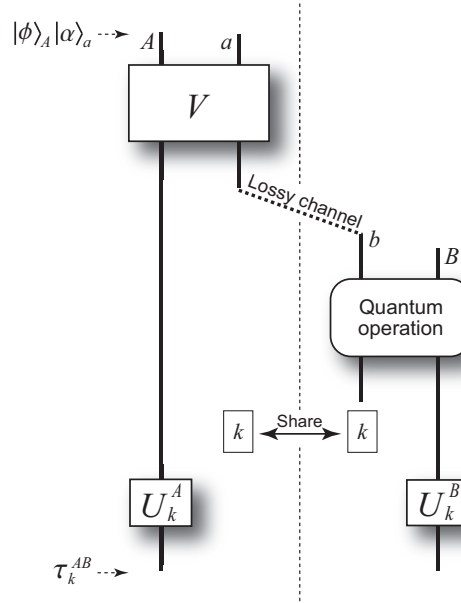}
  \end{center}
  \caption{The scenario of entanglement generation protocols. $\ket{\phi}_A:=\sum_{j=0,1} \sqrt{q_j} e^{i \Theta_j} \ket{j}_A$. Bob's quantum operation returns qubit $B$ in state depending on outcome $k$, and he shares the outcome with Alice by using classical communication. 
   }
  \label{fig1}
\end{figure}

To generate such an entangled state, Alice and Bob execute the following steps (Fig.~\ref{fig1}):
(i) Alice prepares qubit $A$ in her desired state $\ket{\phi}_A= \sum_{j=0,1} e^{i \Theta_j} \sqrt{q_j} \ket{j}_A $ with real parameters $\Theta_j$, $q_j \ge 0$, and $\sum_j q_j=1$, and she makes it interact with a pulse in a coherent state $\ket{\alpha}_a=e^{-|\alpha|^2/2} e^{\alpha \hat{a}^\dag} \ket{0}_a$ via a unitary operation $\hat{V}$ of Eq.~(\ref{eq:V}).
(ii) Alice sends the pulse $a$ to Bob, through a lossy channel described by an isometry
\begin{equation}
\hat{N} \ket{\alpha}_a = \ket{ \sqrt{T} \alpha}_b \ket{\sqrt{1-T} \alpha}_{E},
\end{equation}
where $0<T<1$ is the transmittance of the channel and system $E$ is the environment.
(iii) Upon receiving the pulse in mode $b$, 
Bob may perform arbitrary operations and measurements involving pulse $b$ and his memory qubit $B$, and declare success outcome $k$ occurring with a probability $p_k$ or failure.
(iv) If Step (iii) succeeds, depending on the outcome $k$, Alice and Bob apply a local unitary operation $\hat{U}^A_k \otimes \hat{U}^B_k$ to the obtained state, in order to satisfy that the final state $ \hat{\tau}_k^{AB}$ is contained in the subspace spanned by $\{\ket{\Phi^{\pm}}_{AB}\}$, and also that the nearest Bell state to the state $ \hat{\tau}_k^{AB}$ is $\ket{\Phi^+}_{AB}$.

We evaluate the performance of the protocols by the total success probability,
\begin{equation}
P_s=\sum_k p_k,
\end{equation}
and the averaged fidelity of the obtained entangled states
\begin{equation}
F= \frac{1}{P_s} \sum_k p_k F_k,
\end{equation}
where $F_k$ is 
\begin{equation}
F_k:= \bra{\Phi^+}  \hat{\tau}_k^{AB}  \ket{\Phi^+}.
\end{equation}
Thanks to the choice of the unitary operation in Step (iv), $F_k$ is equivalent to so-called {\em singlet fraction} \cite{B96a}.
Since $\hat{\tau}_k^{AB}$ is contained in the subspace spanned by $\{\ket{\Phi^{\pm}}_{AB}\}$, $F_k \ge 1/2$ holds. This means
\begin{equation}
F\ge 1/2. \label{F1/2}
\end{equation}

We also allow Alice and Bob to switch among two or more protocols probabilistically.
The performance of such a mixed protocol is determined as follows.
Suppose that Alice and Bob can execute a protocol with performance $(P_s^{(1)},F^{(1)})$ and a protocol with performance $(P_s^{(2)},F^{(2)})$.
Then, by choosing these protocols with probabilities $\{r,1-r\}$,
Alice and Bob can achieve performance $(P_s',F')$ determined by
\begin{equation}
\left(
  \begin{array}{c}
    P_s'   \\
    P_s' F'   \\
  \end{array}
\right)
= r \left(
  \begin{array}{c}
    P_s^{(1)}   \\
    P_s^{(1)} F^{(1)}   \\
  \end{array}
\right) 
+ (1-r) \left(
  \begin{array}{c}
     P_s^{(2)}   \\
    P_s^{(2)} F^{(2)}   \\
  \end{array}
\right). \label{conv}
\end{equation} 
It is thus convenient to describe the performance of a protocol by point $(P_s,P_s F)$.
Then, the set of achievable points $(P_s,P_s F)$ forms a convex set.

\section{An upper bound on the performance of a single-error-type entanglement generation protocol}\label{se:3}

We first introduce a protocol equivalent to the single-error-type entanglement generation protocol.
Steps (i) and (ii) indicate that, when the pulse arrives at Bob, the state of the total system $A b E$ is written in the form of
\begin{equation}
 \ket{\psi}_{A b E} = \sum_{j=0,1} \sqrt{q_j} \ket{j}_A \ket{u_j}_{b}
  \ket{v_j}_E 
\label{eq:abe}
\end{equation}
with $0 \le q_0 \le 1 $, $q_0+q_1=1$, and 
\begin{equation}
 |\braket{u_1}{u_0}|^{1-T} = |\braket{v_1}{v_0}|^T>0.
 \label{tra}
\end{equation}

\begin{figure}[b]
  \begin{center}
    \includegraphics[keepaspectratio=true,height=80mm]{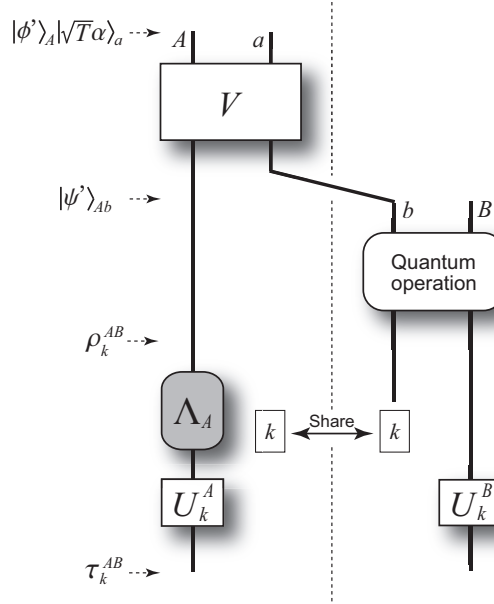}
  \end{center}
  \caption{An imaginary protocol equivalent to the real protocol in Fig.~1. $\ket{\phi'}_A:=\sum_{j=0,1} \sqrt{q_j} e^{i \Theta_j+ i (-1)^j \varphi} \ket{j}_A$. Channel $a \to b$ becomes ideal at the expense of the application of a phase-flip channel $\Lambda_A$.
  }
  \label{fig2}
\end{figure}

Let us define a phase flip channel $\Lambda_A$ on qubit $A$ by
\begin{equation}
\Lambda_A(\hat{\rho}):=f \hat{\rho} + (1-f) \hat{\sigma}_z^A \hat{\rho}
\hat{\sigma}_z^A \label{Lambda}
\end{equation}
with
\begin{equation}
f:=\frac{1+|\braket{v_1}{v_0}|}{2}=\frac{1+|\braket{u_1}{u_0}|^{\frac{1-T}{T}}}{2} \label{fdef}
\end{equation}
and
$\hat{\sigma}_z^A := \ket{0} \bra{0}_{A} -\ket{1} \bra{1}_{A}$.
From Eqs.~(\ref{eq:abe}), (\ref{Lambda}), and (\ref{fdef}), we have 
\begin{equation}
{\rm Tr}_{E} [ \ket{\psi}\bra{\psi}_{A b E} ] = \Lambda_A(\ket{\psi'} \bra{\psi'}_{A b}),
\end{equation}
where 
\begin{equation}
\ket{\psi'}_{A b}:= \sum_{j=0,1}  \sqrt{q_j} e^{i (-1)^j  \varphi } \ket{j}_A \ket{u_j}_{b} \label{eq:psi'} 
\end{equation}
with $2 \varphi:=\arg [\braket{v_1}{v_0}]$. 
The effect of the lossy channel is thus equivalently described as
preparation of $\ket{\psi'}_{A b}$ followed by
$\Lambda_A$. Since any operation of Bob commutes with $\Lambda_A$, 
the protocol is equivalent to the following sequence (Fig.~\ref{fig2}):
(1) System $A b$ is prepared in $\ket{\psi'}_{A b}$;
(2) Bob's successful measurement leaves system $AB$
in a state $\hat{\rho}^{AB}_k$;
(3) $\Lambda_A$ is applied on qubit $A$.

In what follows, according to the equivalent protocol of Fig.~\ref{fig2},
we show that, for fixed $T$ and $|\braket{u_1}{u_0}|$, the performance $(P_s,P_s F)$ of an arbitrary protocol must be in the triangle with the apexes,
\begin{equation}
\eqalign{
X_0:=(0,0), \cr
X_1:=\left( 1-|\braket{u_1}{u_0}|, (1-|\braket{u_1}{u_0}|) \frac{1+|\braket{u_1}{u_0}|^{\frac{1-T}{T}}}{2} \right), \cr
X_2:= \left(1,1/2 \right). }\label{optre}
\end{equation}

a) {\em $|q_0-q_1|= 1$ or $|\braket{u_1}{u_0}|=1$}. In these cases, from Eq.~(\ref{eq:psi'}), $\ket{\psi'}_{Ab}$ is a product state between system $A$ and $b$.
This implies that $\hat{\tau}_k^{AB}$ is a separable state, which means $F_k \le 1/2$. From Eq.~(\ref{F1/2}), $F=1/2$.
 Thus, in this case, the performance $(P_s, P_s F)$ of protocols must be on the segment $X_0 X_2$. 

b) {\em  $|q_0-q_1| < 1$ and $|\braket{u_1}{u_0}|<1$}. 
As stated in Step (iv), whenever Bob declares success outcome $k$, the state $\hat{\tau}_k^{AB}$ of their qubits satisfies 
\begin{equation}
 \bra{\Psi^\pm} \hat{\tau}_k^{AB} \ket{\Psi^\pm} = \bra{ \Psi'^{\pm}_k } \Lambda_A ( \hat{\rho}_k^{AB}) \ket{ \Psi'^{\pm}_k} =0 \label{eq:one-type-error}
\end{equation}
with $\ket{ \Psi'^{\pm}_k }_{AB} := \hat{U}^{A \dag}_k \otimes \hat{U}^{B \dag}_k \ket{ \Psi^{\pm}}_{AB}=(\ket{x_k^0}_A \ket{y_k^1}_B \pm \ket{x_k^1}_A \ket{y_k^0}_B)/\sqrt{2}$, $\ket{ \Psi^{\pm}}_{AB}:=(\ket{01}_{AB} \pm \ket{10}_{AB})/\sqrt{2}$, $\ket{x^j_k}_A:=\hat{U}^{A \dag}_k \ket{j}_A$, and $\ket{y^j_k}_B:=\hat{U}^{B \dag}_k \ket{j}_B$ $(j=0,1)$.
Since $\hat{\rho}^{AB}_k$ is positive and $0<f<1$, 
Eq.~(\ref{eq:one-type-error}) indicates
\begin{eqnarray}
\sqrt{\hat{\rho}^{AB}_k} \ket{ \Psi'^{\pm}_k}_{AB}=0, \label{ortho1} \\
\sqrt{\hat{\rho}^{AB}_k} \hat{\sigma}_z^A \ket{ \Psi'^{\pm}_k }_{AB}=0, \label{ortho2}
\end{eqnarray}
for both $\pm$.
Note that Eq.~(\ref{ortho1}) implies
\begin{equation}
\eqalign{
\hat{\rho}^{AB}_k=&  \frac{1+a_k}{2}  \ket{ \Phi'^{+}_k}\bra{ \Phi'^{+}_k}_{AB} + \frac{1-a_k}{2}  \ket{ \Phi'^{-}_k}\bra{ \Phi'^{-}_k}_{AB} \cr 
& + \frac{b_k}{2} \ket{ \Phi'^{+}_k}\bra{ \Phi'^{-}_k}_{AB} + \frac{b_k^*}{2} \ket{ \Phi'^{-}_k}\bra{ \Phi'^{+}_k}_{AB} ,} \label{rho_k}
\end{equation}
where $\ket{ \Phi'^{\pm}_k}_{AB} := \hat{U}^{A \dag}_k \otimes \hat{U}^{B \dag}_k \ket{ \Phi^{\pm} }_{AB}
=(\ket{x^0_k}_A \ket{y^{0}_k}_B\pm \ket{x^1_k}_A \ket{y^{1}_k}_B)/\sqrt{2}$, and the positivity of $\hat{\rho}_k^{AB}$ implies
\begin{equation}
a_k^2 + |b_k|^2 \le 1. \label{a_kb_k}
\end{equation}
Note that $0 \le a_k \le 1$ is satisfied by the choice of the unitary operation $\hat{U}^{A }_k \otimes \hat{U}^{B }_k$ in Step (iv).
Adding and subtracting Eqs. (\ref{ortho1}) and (\ref{ortho2}), we obtain 
\begin{eqnarray}
\eqalign{\sqrt{ \hat{\rho}^{AB}_k}  \ket{x^0_k}_A \ket{y^1_k}_B = \sqrt{ \hat{\rho}^{AB}_k} \hat{\sigma}_z^A \ket{x^0_k}_A \ket{y^{ 1}_k}_B \cr
=\sqrt{ \hat{\rho}^{AB}_k}  \ket{x^1_k}_A \ket{y^{0}_k}_B = \sqrt{ \hat{\rho}^{AB}_k} \hat{\sigma}_z^A \ket{x^1_k}_A \ket{y^{0}_k}_B  =0.}
\label{zero}
\end{eqnarray}
Since $\hat{\rho}^{AB}_k\neq 0$, 
the four states, $\ket{x^0_k}_A \ket{y^{1}_k}_B$, 
$\hat{\sigma}_z^A \ket{x^0_k}_A \ket{y^{1}_k}_B$,
$\ket{x^1_k}_A \ket{y^{0}_k}_B$, and $\hat{\sigma}_z^A
\ket{x^1_k}_A \ket{y^{0}_k}_B $, 
must be linearly dependent,
which only happens when
$\{\ket{x^j_k}_A\}_{j=0,1}$ is a set of eigenvectors of $\hat{\sigma}_z^A$.
Combining this fact with Eq.~(\ref{rho_k}), we obtain
\begin{equation}
\hat{\rho}_k^{A}:= {\rm Tr}_B [ \hat{\rho}_k^{AB}] = \frac{\hat{1}^A +z_k \hat{\sigma}_z^A }{2}, \label{rhoA_k}
\end{equation}
where $z_k:=\pm {\rm Re} (b_k)$.

The fidelity $F_k$ of the final state is given by
$F_k= \bra{\Phi^+} \hat{\tau}_k^{AB} \ket{\Phi^+} =\bra{ \Phi'^{+}_k }\Lambda_A (\hat{\rho}_k^{AB}) \ket{ \Phi'^{+}_k }$.
Since $\{\ket{x^j_k}_A \}_{j=0,1}$ is an eigenbasis of
$\hat{\sigma}_z^A$, we have 
$\hat{\sigma}_z^A\ket{ \Phi'^{+}_k }=\pm \ket{ \Phi'^{-}_k }$, which means $F_k=f \bra{ \Phi'^{+}_k }\hat{\rho}^{AB}_k \ket{ \Phi'^{+}_k }
+(1-f)\bra{ \Phi'^{-}_k }\hat{\rho}^{AB}_k \ket{ \Phi'^{-}_k}$.
From Eqs.~(\ref{rho_k}) and (\ref{fdef}), the fidelity $F_k$ is rewritten as
\begin{equation}
F_k= \frac{1}{2} (1+ |\braket{v_1}{v_0}| a_k).\label{F_k}
\end{equation}
Combining this equation, Eq.~(\ref{a_kb_k}), and the definition of $z_k$, we have
\begin{equation}
 \left( \frac{2F_k-1}{|\braket{v_1}{v_0}|} \right)^2 +z_k^2  \le 1. \label{traF}
\end{equation}

Let us consider the success probability of the protocol.
Suppose that Bob's failure measurement returns a state $\hat{\rho}_f^{AB}$ with probability $1-P_s$. Since Alice does nothing until the end of Bob's generalized measurement, Alice's averaged density operator is unchanged through the measurement, i.e.,
\begin{equation}
\hat{\psi}'^A = P_s \hat{\rho}^{A}_s + (1-P_s)  \hat{\rho_f}^{A}, \label{psi'-eq}
\end{equation}
where $\hat{\psi}'^A :={\rm Tr}_b [\ket{\psi'} \bra{\psi'}_{Ab}]$, $\hat{\rho}^{A}_s:=  (\sum_k p_k \hat{\rho}^{A}_k )/P_s$ and $\hat{\rho_f}^{A}:={\rm Tr}_B [\hat{\rho_f}^{AB}]$. 
Eq.~(\ref{eq:psi'}) indicates that $\hat{\psi}'^A$ is in the form of
\begin{equation}
\hat{\psi}'^A = \frac{\hat{1}^A + x_0 \hat{\sigma}_x^A + y_0 \hat{\sigma}_y^A + z_0 \hat{\sigma}_z^A}{2},
\end{equation}
where $\hat{\sigma}_x^A:=\ket{0}\bra{1}_A+\ket{1}\bra{0}_A$, $\hat{\sigma}_y^A:= -i \ket{0}\bra{1}_A+i\ket{1}\bra{0}_A$, and $x_0$, $y_0$ and $z_0$ satisfy
\begin{eqnarray}
z_0=q_0-q_1, \\
x_0^2+y_0^2 = 4 q_0 q_1 |\braket{u_1}{u_0}|^2= (1-z_0^2)  |\braket{u_1}{u_0}|^2. \label{x_0}
\end{eqnarray}
On the other hand, $\hat{\rho}_s^A$ is written as
\begin{equation}
\hat{\rho}_s^A = \frac{1}{P_s} \sum_k p_k \hat{\rho}_k^A= \frac{\hat{1}+z_s \hat{\sigma}_z^A}{2},
\end{equation}
where $z_s:=(\sum_{k} p_k z_k)/P_s$, and it satisfies 
\begin{equation}
\left( \frac{2F-1}{|\braket{v_1}{v_0}|} \right)^2 +z_s^2 \le 1 \label{traFz}
\end{equation}
from Eq.~(\ref{traF}) and the convexity of function $x^2$. 
Note that this inequality implies
\begin{equation}
F \le \frac{1+|\braket{v_1}{v_0}|}{2}=\frac{1+|\braket{u_1}{u_0}|^{\frac{1-T}{T}}}{2}, \label{triF}
\end{equation}
where we used Eq.~(\ref{tra}).
We also decompose $\hat{\rho}_f^A$ as
\begin{equation}
\hat{\rho}_f^A = \frac{\hat{1}^A + x_f \hat{\sigma}_x^A + y_f \hat{\sigma}_y^A + z_f \hat{\sigma}_z^A}{2}
\end{equation}
with real numbers $x_f,y_f,z_f$ satisfying
\begin{equation}
x_f^2+y_f^2+z_f^2 \le 1. \label{eq:normali}
\end{equation}
From Eq.~(\ref{psi'-eq}), we have
\begin{eqnarray}
\eqalign{ 
x_0 =  (1-P_s) x_f, \cr 
y_0  =  (1-P_s) y_f, \cr
z_0  = P_s z_s +  (1-P_s) z_f.} 
\end{eqnarray}
From these equations, Eq.~(\ref{x_0}) and Eq.~(\ref{eq:normali}),
we obtain
\begin{equation}
g(P_s):=P_s^2 (1-z_s^2) - 2  P_s (1-z_0 z_s)  
+ (1- |\braket{u_1}{u_0}|^2) (1-z_0^2)  
\ge 0, \label{eq:g}
\end{equation}
or equivalently, we have 
\begin{equation}
\eqalign{ \left[ (1- |\braket{u_1}{u_0}|^2) z_0 -P_s z_s \right]^2 
\le 
 [1-(1-z_s^2) |\braket{u_1}{u_0}|^2 ] \cr  
\times \left( P_s-\frac{1-|\braket{u_1}{u_0}|^2}{1-|\braket{u_1}{u_0}| \sqrt{1-z_s^2}} \right) 
 \left( P_s-\frac{1-|\braket{u_1}{u_0}|^2}{1+|\braket{u_1}{u_0}| \sqrt{1-z_s^2}} \right). } \label{eq:g'}
\end{equation}
Since $z_0^2<1$ and $0<|\braket{u_1}{u_0}|<1$, we have
\begin{eqnarray}
g(1-|\braket{u_1}{u_0}|^2)&=-\left(1-|\braket{u_1}{u_0}|^2\right) [
   \left(1-z_s^2\right) |\braket{u_1}{u_0}|^2 
   +(z_0-z_s)^2 ]  \\
   &<0, 
\end{eqnarray}
and
\begin{equation}
g(1)= - \left(1-z_0^2\right)|\braket{u_1}{u_0}|^2 -(z_0-z_s)^2<0,
\end{equation}
which mean $g(P_s)<0$ for $P_s \ge 1-|\braket{u_1}{u_0}|^2$ because $g(P_s)$ is linear or convex. 
Thus, Eq.~(\ref{eq:g}) implies 
\begin{equation}
P_s <  1- |\braket{u_1}{u_0}|^2. \label{up}
\end{equation}
To satisfy inequality (\ref{eq:g'}), the right-hand side of the inequality should be nonnegative, which occurs only when
\begin{equation}
P_s \le  \frac{1-|\braket{u_1}{u_0}|^2}{1+|\braket{u_1}{u_0}| \sqrt{1-z_s^2}}
\end{equation}
under the condition of Eq.~(\ref{up}).
Combining Eq.~(\ref{traFz}), we have
\begin{equation}
P_s \le \frac{1-|\braket{u_1}{u_0}|^2}{1+|\braket{u_1}{u_0}| \left( \frac{2F-1}{|\braket{v_1}{v_0}|} \right)}, \label{Ps-up}
\end{equation}
which can be rewritten as
\begin{eqnarray}
P_s F &\le& \frac{1}{2} \left( 1- \frac{|\braket{v_1}{v_0}|}{|\braket{u_1}{u_0}|} \right) P_s 
 + \frac{1}{2} (1-|\braket{u_1}{u_0}|^2)  \frac{|\braket{v_1}{v_0}|}{|\braket{u_1}{u_0}|} \\
&=& \frac{1}{2} \left( 1- |\braket{u_1}{u_0}|^{\frac{1-2T}{T}} \right) P_s 
 + \frac{1}{2} (1-|\braket{u_1}{u_0}|^2)  |\braket{u_1}{u_0}|^{\frac{1-2T}{T}},
\label{traPF}
\end{eqnarray}
where we used Eq.~(\ref{tra}).

Since Eq.~(\ref{F1/2}), Eq.~(\ref{triF}), and Eq.~(\ref{traPF}) must be satisfied at the same time, the performance $(P_s, P_s F)$ of an arbitrary protocol must be in the triangle with the apexes $X_0$, $X_1$, and
\begin{equation}
X_3:= \left(1-|\braket{u_1}{u_0}|^2, \frac{1}{2}(1-|\braket{u_1}{u_0}|^2)\right), 
\end{equation}
which is included in the triangle $X_0X_1X_2$.
This completes the proof.

\section{Simulatability of an arbitrary protocol via symmetric protocols}\label{se:4}

Here we show that the performance of an arbitrary protocol, which is in the triangle defined by Eq.~(\ref{optre}) with fixed $T$ and $|
\braket{u_1}{u_0}|$, is simulatable by utilizing a protocol in Ref.~\cite{S08}.
In the protocol \cite{S08}, Alice starts with preparing system $A$ in a symmetric state $  \ket{\phi}_A = (\ket{0}_A+\ket{1}_A)/\sqrt{2}$, and, upon receiving pulses from Alice, Bob carries out a measurement that is composed of a simple combination of linear optical elements and photon-number-resolving detectors. 
Let us call it symmetric protocol in what follows.
With a proper choice of the intensity of pulse $a$, the symmetric protocol can achieve $(P_s, P_s F)$ with
\begin{eqnarray}
\eqalign{
P_s = 1 - u, \cr
F= \frac{1+u^{\frac{1-T}{T}}}{2},} \label{symp}
\end{eqnarray}
for any $u$ with $0 < u \le 1$ \cite{S08}.
This indicates that the symmetric protocol can achieve performances $(P_s, P_s F)=X_0$ by choosing $u=1$, and $(P_s, P_s F)=X_1$ by choosing $u=|\braket{u_1}{u_0}|$.
On the other hand, the performance $(P_s, P_s F)=X_2$ is also achievable by a trivial protocol in which Alice and Bob prepare their memories in state $\ket{00}_{AB}$ and declare success all the time.
The achievability of points $X_0$, $X_1$, and $X_2$ indicates that all the points in the triangle $X_0X_1X_2$ are achievable by mixing.
Since this fact holds for any $|\braket{u_1}{u_0}|$, we conclude that, for given $T$, the performance of an arbitrary protocol is simulatable by combining symmetric protocols and the trivial protocol.

\section{Optimal performance of single-error-type entanglement generation}\label{se:5}

\begin{figure}[tb]
  \begin{center}
    \includegraphics[keepaspectratio=true,height=45mm]{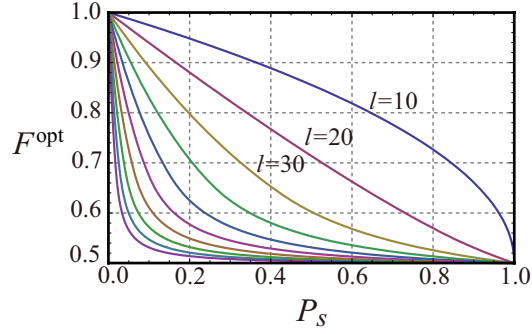}
  \end{center}
  \caption{The optimal performances of single-error-type entanglement generation for $10 \le l \le 100$ km at intervals of $10$ km, where we assume $T=e^{-l/l_0}$ and $l_0=25$ km (corresponding to $ \sim 0.17$ 
dB/km attenuation). }
  \label{fig:azuma4-6-2.eps}
\end{figure}

Here we calculate the optimal performance of the mixture of arbitrary single-error-type entanglement generation protocols for given $T$.
As shown in the preceding section, for any $T$, the performance $(P_s, P_s F)$ of an arbitrary protocol is achievable by mixing symmetric protocols and the trivial protocol.
Since the performance achieved by a symmetric protocol or the trivial protocol can be described by a point $(P_s,P_s F)=(P_s, P_s F^{\rm sym} (P_s) )$ with 
\begin{equation}
F^{\rm sym} (P_s):=\frac{1+(1-P_s)^{\frac{1-T}{T}}}{2} ,\; (0 \le P_s \le 1), \label{xi}
\end{equation}
the performance of the mixture of arbitrary protocols must be in the convex hull of the region ${\cal S}:=\{ (P_s, P_s F) \;|\; 0 \le P_s \le 1,\; 1/2 \le F \le F^{\rm sym}(P_s) \}$.
In what follows, we show that the convex hull, ${\rm Conv}({\cal S})$, is given by the region ${\cal C_S}:=\{(P_s, P_s F)\;|\; 0 \le P_s \le 1,\;1/2 \le F \le F^{\rm opt} (P_s) \}$ with $F^{\rm opt} (P_s)$ defined by
\begin{equation}
F^{\rm opt} (P_s) :=
\cases{  \frac{1+(1-P_s)^{\frac{1-T}{T}}}{2},  &  $ (P_s \le \frac{T}{1-T}), $ \\
     \frac{1}{2} + \frac{1-P_s}{2 P_s} \frac{T}{1-2T} \left( \frac{1-2T}{1-T} \right)^{\frac{1-T}{T}},  & $ (P_s > \frac{T}{1-T}). $ \\ } \label{opt}
\end{equation}
Note that $P_s > T/(1-T)$ holds only when $T<1/2$.
The tight upper bound $F^{\rm opt} (P_s)$ is depicted in Fig.~\ref{fig:azuma4-6-2.eps}.

Let us proceed to the proof of ${\cal C}_{\cal S}= {\rm Conv} ({\cal S})$. From Eq.~(\ref{xi}), we have
\begin{eqnarray}
\eqalign{
\frac{{\rm d}   P_s F^{\rm sym}(P_s) }{{\rm d} P_s} =&\frac{1}{2} \left[ 1+ 
\left(1-\frac{P_s}{T} \right) (1-P_s)^{\frac{1-2T}{T}} \right], \label{1bibun}\cr
\frac{{\rm d}^2 P_s F^{\rm sym}(P_s) }{{\rm d} {P_s}^2}=& \frac{1}{2} \frac{1-T}{T} \left( \frac{P_s}{T} - 2 \right) (1-P_s)^{\frac{1-3T}{T}}.}
\end{eqnarray}
The latter equation indicates
\begin{equation}
\eqalign{ \frac{{\rm d}^2 P_s F^{\rm sym}(P_s) }{{\rm d} {P_s}^2}>&0,\;   (P_s > 2T), \label{final}   \cr
\frac{{\rm d}^2 P_s F^{\rm sym}(P_s) }{{\rm d} {P_s}^2}\le &0,\;   (P_s \le 2T).}  
\end{equation}

a) {\rm $T \ge 1/2$}. In this case, $F^{\rm opt}(P_s) =F^{\rm sym} (P_s)$, and hence ${\cal S}={\cal C}_{\cal S}$. In addition, Eq.~(\ref{final}) indicates that $P_s F^{\rm sym}(P_s) $ is concave for $0\le P_s \le1$. These facts imply that ${\rm Conv}({\cal S})$ is equivalent to ${\cal S}$, namely, to ${\cal C_S}$.

b) {\rm $T < 1/2$}. Let $P_s^*$ be $P_s^*:=T/(1-T)$.
The proof begins with noting the following facts:
(i) $F^{\rm opt}(P_s)=F^{\rm sym}(P_s)$ for $0 \le P_s < P_s^*$;
(ii) $F^{\rm opt}(P_s^*)=F^{\rm sym}(P_s^*)$;
(iii) $F^{\rm opt}(1)=F^{\rm sym}(1)$;
(iv) $P_s F^{\rm opt} (P_s)$ and $({\rm d} P_s F^{\rm opt} (P_s))/ ({\rm d} P_s) $ are continuous at $P_s=P_s^*$;
(v) 
\begin{equation}
\frac{{\rm d}^2 P_s F^{\rm opt} (P_s)}{{\rm d} {P_s}^2}
\cases{ <0, & $ (0 \le P_s < P_s^*),$ \\
    =0, & $ (P_s^* <  P_s);$ \\}
\label{dd}
\end{equation}
(vi) $F^{\rm opt} (P_s) > F^{\rm sym } (P_s)$ for $P_s^* < P_s <1$.
Facts (i)-(v) are easily confirmed from Eqs.~(\ref{xi})-(\ref{opt}).
Fact (vi) is proven by facts (ii)-(iii), 
\begin{equation}
\frac{{\rm d} P_s F^{\rm opt} (P_s^*) }{{\rm d} P_s} =  \frac{{\rm d} P_s F^{\rm sym} (P_s^*) }{{\rm d} P_s}, 
\end{equation}
and by Eqs.~(\ref{final})-(\ref{dd}). Facts (iv)-(v) show that ${\cal C}_{\cal S}$ is convex.
Facts (i)-(iii) and (vi) imply $ {\cal S} \subset {\cal C}_{\cal S}$.
From facts (i)-(v), we have ${\cal C}_{\cal S} \subset {\rm Conv}({\cal S})$.
Therefore, we conclude ${\rm Conv}({\cal S})={\cal C}_{\cal S}$.

\section{Summary}\label{se:6}
In conclusion, we have provided the tight upper bound on the performances of protocols that generate entanglement with only one type of error by transmitting pulses in coherent states through a lossy channel.
As represented by Eq.~(\ref{opt}), the tight upper bound is stated in terms of the success probability $P_s$ and the average singlet fraction $F$ of generated entanglement,
and is determined only by the transmittance $T$ of the channel.
In addition, we have shown that the upper bound is achievable without large-scale quantum operations, namely by utilizing a simple protocol \cite{S08} composed of linear optical elements and photon-number-resolving detectors.

The arts enabling us to derive such a general bound can be summarized as follows. The proof begins with replacing the real protocol in Fig.~\ref{fig1} by an equivalent (virtual) protocol in Fig.~\ref{fig2}. Thanks to the replacement, the effect of the optical loss in the practical channel is reduced to a {\em local} phase-flip channel acting on Alice's memory, and the quality of final entanglement is bounded by the form of the local density operator of the memory $A$ fed to the phase-flip channel (see Eqs.~(\ref{rhoA_k}) and (\ref{traF})).
Since the local density operator can only be altered by Bob remotely at the expense of a failure probability, 
we are led to Eq.~(\ref{psi'-eq}) relating the change in the Alice's local density operator and the success probability. 
This relation enables us to derive a trade-off relation Eq.~(\ref{traPF}) between the success probability $P_s$ and the average singlet fraction $F$, which leads to the tight upper bound of arbitrary protocols.

Throughout this paper, we have focused on the entanglement generation
protocols with only one type of error, based on the fact that the
known simple distillation protocols work more efficiently against
such a restricted type of errors. This has allowed us to treat the
entanglement generation protocols separately from distillation
protocols. If we look into the properties of the distillation
protocols in more detail, there is a possibility that accepting
multiple types of errors for higher success probability in the
generation protocol could lead to a better result if there exists a
distillation protocol with a less penalty on the multiple types of
errors. Pursuing such a possibility is important for implementation
of quantum repeaters, and is also interesting in connection to the
fundamental question of what is the best way of distributing
entanglement against an optical loss in the channel. We expect that
the arts introduced here may be also useful in solving such general
problems in the search of good entanglement generation protocols in
hybrid quantum repeaters. 

\section*{Acknowledgement}
We would like to thank \c{S}ahin Kaya \"Ozdemir, Ryo Namiki, Takashi Yamamoto, and Hitoshi Takeda for 
valuable discussions. We acknowledge the support
of a MEXT Grant-in-Aid for Scientific Research on
Innovative Areas 21102008, a MEXT Grant-in-Aid for
the Global COE Program, and JSPS Grant-in-Aid for Scientific
Research (C) 20540389. K.A. is
supported by JSPS Research Fellowships for Young Scientists.


\end{document}